\documentstyle[aps,preprint,prl,citesort,epsf]{revtex}

\tightenlines

\newcommand{\eqref}[1]{Eq.~(\protect\ref{#1})}
\newcommand{\figref}[1]{Fig.~\protect\ref{#1}}

\begin{document}

\draft

\title{Escape Time Weighting of Unstable Stationary Solutions of 
Spatiotemporal Chaos}

\author{Scott M. Zoldi$^{\dagger}$}


\address{$^{\dagger}$ Theoretical Division and Center for 
Nonlinear Studies, Los Alamos National Laboratory, 
Los Alamos, New Mexico 87545}

\date{\today}
\maketitle


\begin{abstract}
By computing 254 unstable stationary solutions of the
Kuramoto-Sivashinsky equation in the extensive chaos 
regime (Lyapunov fractal dimension~$D=8.8$), we 
find that 30\% satisfy the symmetry of the time-average 
pattern of the spatiotemporal chaos.  Using a symmetry 
pruning of unstable stationary solutions, the escape-time 
weighting average converges to the time-average 
pattern of the chaotic attractor as $O(1/N)$, where $N$ 
is the total number of unstable stationary solutions 
and unstable periodic orbits in the average. 
\end{abstract}


           

\narrowtext

A topic of great current interest in spatiotemporal chaos theory is to
determine the appropriate quantities to characterize and distinguish
chaotic states from one another.  Attempts to distinguish states using
dimension densities~\cite{Egolf95prl,OHern96,Zoldi97prl}, correlation
lengths~\cite{Tufillaro89,Morris93,Hu95prl2}, and pattern dependent
order-parameters~\cite{Gunaratne95,Mecke94} reduce the dynamics to
scalar quantities with varying degrees of sensitivity to parameter
changes and often different parametric dependencies.  Various
projections of the chaotic dynamics have proven more sensitive to
parameter changes or inhomogeneities in the dynamics.  These have included
histograms of the intensity of multi-mode lasers~\cite{Roy91}, spatial
averages of the curvature of a fluid surface in Faraday wave
experiments~\cite{Gluckman95}, and averages of shadowgraph images in
convection experiments~\cite{Ning93prl}.  Recent attempts to predict
the structure in the projections using unstable periodic orbits (UPOs)
in low-dimensional chaos~\cite{Zoldi98prl2} and spatiotemporal
chaos~\cite{Zoldi98pre} have demonstrated that the escape-time
weighting can approximate averages of chaos based on the moments of a
computed set of UPOs.  This suggests that UPOs may be more appropriate
to characterize chaos because an infinity of order-parameters
(including those discussed above) can be approximated using a small
set of UPOs.  Also, the parametric dependence of the order-parameters
are easily determined by using Newton continuation methods to evaluate
the set of UPOs in nearby parameter regimes.

The utility of UPOs to characterize chaos is based on the concept that
Axiom-A chaotic systems have a dense set of UPOs associated with the
strange attractor~\cite{Smale67}.  In Axiom-A chaotic attractors with
a symbolic dynamics, trace-formulas of a small number of {\em ordered}
UPOs can be used to estimate dynamical invariants and averages of
chaos~\cite{Cvitanovic95,Bauer97}.  Unfortunately, most spatiotemporal
chaotic systems are not Axiom-A~\cite{Dawson94} and do not have a
symbolic dynamics which make trace formulas inappropriate for
high-dimensional chaos.  However, the escape-time weighting (ETW) of
UPOs has been used to approximate averages in high-dimensional
chaos~\cite{Zoldi98pre} where trace-formulas often fail primarily due
to a loss of symbolic dynamics~\cite{Zoldi98prl2}.  To date, the ETW
average of the easier to compute unstable stationary solutions of the
spatiotemporal chaos has not been evaluated.

Our goal in this report is to demonstrate that the convergence
of the ETW average of UPOs in the spatiotemporal chaotic
Kuramoto-Sivashinsky equation can be improved using unstable
stationary solutions (USS).  We demonstrate that ETW converges as
$O(1/N)$ with $N$ computed UPOs and USS.  As no method exists 
to enumerate UPOs and USS in nonlinear partial differential equations, we use 
utilize damped-Newton algorithms~\cite{Zoldi98pre} to compute the 
UPOs and USS.

The unstable stationary solutions computed using damped-Newton methods 
can be isolated from the
chaotic attractor.  Past work computing USS in the
Kuramoto-Sivashinsky equation with periodic boundary conditions
demonstrated that the USS consisted of an unexpectedly large class of
stationary solutions, which include laminar states, N-cell states
(periodic patterns), long-wave modulated N-cell states, giant states
(large amplitude solutions), and traveling wave
states~\cite{Hyman86,Michelson86,Greene88}. Many of these stationary
states were not observed in the chaotic dynamics of the
Kuramoto-Sivashinsky 
equation,
whereas the time-dependent UPOs~\cite{Hyman86,Michelson86,Christiansen97}
did qualitatively resemble the chaotic dynamics.

Motivated by cycle-expansions~\cite{Cvitanovic95,Bauer97}, researchers
became interested in approximating moments of sustained chaos in the
Kuramoto-Sivashinsky equation using UPOs on the chaotic
attractor~\cite{Christiansen97,Zoldi98pre}.  Christiansen,
Cvitanovi\'c, and Putkaradze considered the Kuramoto-Sivashinsky
equation with periodic boundary conditions~\cite{Christiansen97}.
Cycle-expansions of the UPOs were evaluated as the small fractal
dimension~$D<2.1$ of the system allowed an understanding of the
symbolic dynamics of the UPOs.  We consider extensive chaos in the
Kuramoto-Sivashinsky equation (fractal-dimension~$D=8.8$) with {\em
rigid boundary conditions} where the lack of a symbolic dynamics makes
cycle expansions inapplicable.  However, the escape-time weighting of
UPOs~\cite{Zoldi98pre} can be used to approximate the time-average 
pattern of the spatiotemporal chaos.

To explore how a suitably selected set of USS improves the escape-time
weighting average of UPOs, we compute the stationary solutions of
the 1d Kuramoto-Sivashinsky (KS) equation,
\begin{equation}
  \partial_t u = - u \, \partial_x u - \partial_x^2 u - \partial_x^4 u
                    , \qquad x \in [0, L] , \label{ks-eq}
\end{equation}
with `rigid'' boundary conditions $u =\partial_x{u} = 0$ at~$x=0$
and~$x=L$.  To compute stationary solutions, a damped-Newton algorithm
(discussed in Ref~\cite{Zoldi98pre}) is used to find the zeros of the
residual,
\begin{equation}
F= u \, \partial_x u + \partial_x^2 u + \partial_x^4 u,
\end{equation}
which was discretized with~$\triangle x = 0.25$.  The initial guesses
to the damped-Newton algorithm were chosen randomly from the chaotic
attractor and the convergence criteria for finding a USS was
that~$\|\delta u\| < 10^-7\|u_o\|$, where $\delta u$ is the correction
vector, $u_o$ is the solution, and~$\| \cdots \|$ denotes the infinity
norm.  For $L=50$ and with 30,000 initial guesses, we found 254
distinct unstable stationary solutions and no stable solutions.

Fig.~\protect\ref{fig:ks-fixed} shows three of the 254 USS computed by
the damped-Newton algorithm.  Many unstable stationary solutions 
were found to be 
isolated~(Fig.~\protect\ref{fig:ks-fixed}(a) and
Fig.~\protect\ref{fig:ks-fixed}(b)) 
from the chaotic attractor not resembling any chaotic
solution.  Other USS did resemble chaotic solutions
(Fig.~\protect\ref{fig:ks-fixed}(c)) and exhibited small values of the
difference function,
\begin{equation}
G(t) = \|u(t,x) - s(x)\|, 
\end{equation}
where $u(t,x)$ denotes
the spatiotemporal chaotic field, $s(x)$ denotes the stationary
solution, and $\| \cdots \|$ denotes the infinity norm. Unfortunately 
finding the minimum value of~$G(t)$ depends on recurrence times 
of the spatiotemporal chaos, so we prune the set of computed USS by
requiring that stationary solutions have the same symmetry as the
time-average pattern of the chaotic attractor.  We impose the
constraint that the stationary solutions satisfy the condition that
$\partial_x^2 u(0) < 0$ and $\partial_x^2 u(L) > 0$.  This symmetry
pruning results in a set of 80 symmetry related USS.  We note that
this criteria does not remove all isolated USS 
(Fig.~\protect\ref{fig:ks-fixed}(a)).

All of the average patterns of the UPOs computed for the KS equation
in Ref~\cite{Zoldi98pre} satisfy the symmetry pruning defined above
for the USS.  These average patterns were used to
approximate the time-average pattern of the chaotic attractor using
the escape-time weighting.  The escape-time weighting~(ETW),
\begin{equation}
<m> = \frac{\sum_j m_j \tau_j}{\sum \tau_j}.
\end{equation}
approximates an average quantity~$<m>$ of the chaotic attractor in
terms of the same average quantity~$m_j$ for each of the $j$ computed
UPOs, where $\tau=\sum_i 1/\lambda_i$ and $\lambda_i$ are the
nonnegative Lyapunov exponents of the UPOs.  The ETW of 127 UPO
average patterns~$<m>$ qualitatively resembles the time-average
pattern of the attractor~$<u>_T$ (Fig.~\protect\ref{fig:ks-ZZZ}(a)).
To include symmetry pruned USS into the ETW average, the weighting 
becomes~$\tau=\sum_i 1/s_i$, where~$s_i$ are the nonnegative
growth factors of the stationary solution.  Applying ETW to the
spatial profiles of symmetry pruned USS and the UPO average patterns,
we obtain the improved ETW approximation $<m>$ to the average
pattern~$<u>_T$ in Fig.~\protect\ref{fig:ks-ZZZ}(b).  This 
improvement is quantified using the error function~$E(x)$,
\begin{equation}
E(x) = \|<m(x)> - u_T(x)\|,
\end{equation}
where $\| \cdots \|$ is the absolute difference.  The average
error~$<E(x)>$ for the ETW average of both UPOs and USS
(\figref{fig:ks-errorfixandupo}(b)) is 50\% smaller than that of a ETW
average of just USS or a ETW average of just UPOs
(\figref{fig:ks-errorfixandupo}(a)).  Empirically, ETW appears to
converge as $O(1/N)$ with $N$ computed UPOs and USS in a high fractal
dimension chaotic system in agreement with the low-dimensional 
convergence rate of ETW reported in Ref~\cite{Zoldi98prl2}.

In conclusion, of 254 USS computed with a damped-Newton algorithm
approximately 30\% had the same symmetry as the time-average pattern
of the spatiotemporal chaos.  Including the symmetry pruned USS into
the ETW average of the UPOs resulted in an improvement to the average 
consistent with a $O(1/N)$ convergence of ETW with $N$ computed UPOs and USS.  
As ETW does not require the assumptions of Axiom-A dynamics or 
a symbolic dynamics, it constitutes one of the few averaging 
techniques for high fractal dimension chaos.
With improved numerical techniques and computational resources to
compute many UPOs and USS, ETW may provide an important means to
quickly evaluate the parametric dependencies of a variety of 
projections of spatiotemporal chaos.

The author thanks Henry S. Greenside and Joshua Socolar 
for many helpful conversations.  This work was supported in part by
NSF grants NSF-DMS-93-07893 and NSF-CDA-92123483-04, and by DOE grant
DOE-DE-FG05-94ER25214.


\bibliographystyle{prsty}  


\newpage

\begin{figure}   
\caption{Three representative unstable stationary solutions of the
Kuramoto-Sivashinsky equation with~$L=50.0$,~$\triangle x = 0.25$, and
rigid boundary conditions~$u(0)=u(L)=u_x(0)=u_x(L)=0$.  The criterion
for convergence was~$\|\delta u\| < 10^-7 \|u_o\|$, where $\delta u$
is the correction vector, $u_o$ is the solution, and~$\| \cdots \|$
denotes the infinity norm.  {\bf (A)} and {\bf
(B)} are isolated stationary solutions, whereas the stationary 
solution {\bf (C)} has finite
measure on the chaotic attractor.}
\label{fig:ks-fixed}
\end{figure}

\begin{figure}   
\caption {{\bf (A)}:Time-average of the Kuramoto-Sivashinsky chaotic
trajectory~$ \langle u(t,x) \rangle_T$ (solid line) and the
escape-time weighting~$\langle M \rangle$ (dashed line) of 127
unstable periodic orbit time-average patterns.  {\bf (B)}:Time-average of the
Kuramoto-Sivashinsky chaotic trajectory~$\langle u(t,x) \rangle_T$
(solid line) and the escape-time weighting~$\langle M \rangle$ (dashed
line) of 127 unstable periodic orbit time-average patterns 
and 80 symmetry pruned
unstable stationary solutions.}
\label{fig:ks-ZZZ}
\end{figure}

\begin{figure}   
\caption
{{\bf (A)}:~Error~$E(x)$ between the time-average of the
Kuramoto-Sivashinsky chaotic trajectory and the escape-time weighting
of 127 unstable periodic orbit time-average patterns of the
Kuramoto-Sivashinsky equation.  The average error is 0.11.  {\bf
(B)}:Error~$E(x)$ between the time-averaged pattern of the
Kuramoto-Sivashinsky chaotic trajectory and the escape-time weighting
of 80 symmetry pruned unstable stationary solutions and 127 unstable periodic
orbit time-average patterns of the Kuramoto-Sivashinsky equation.  The
average error is 0.05.  }
\label{fig:ks-errorfixandupo}
\end{figure}

\newpage
\centerline{\epsfsize=25.5in \epsfbox{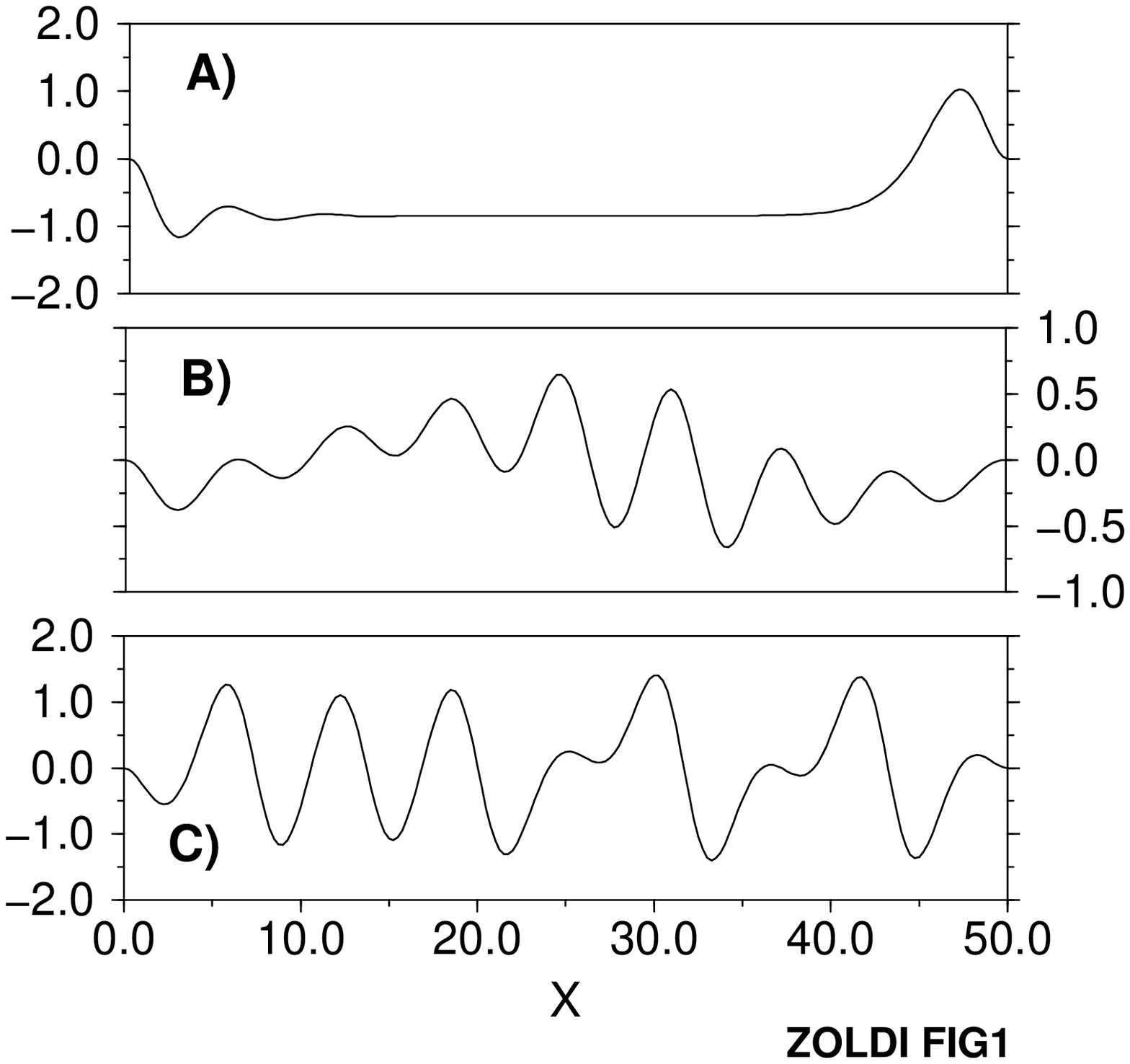}}

\newpage
\centerline{\epsfsize=25.5in \epsfbox{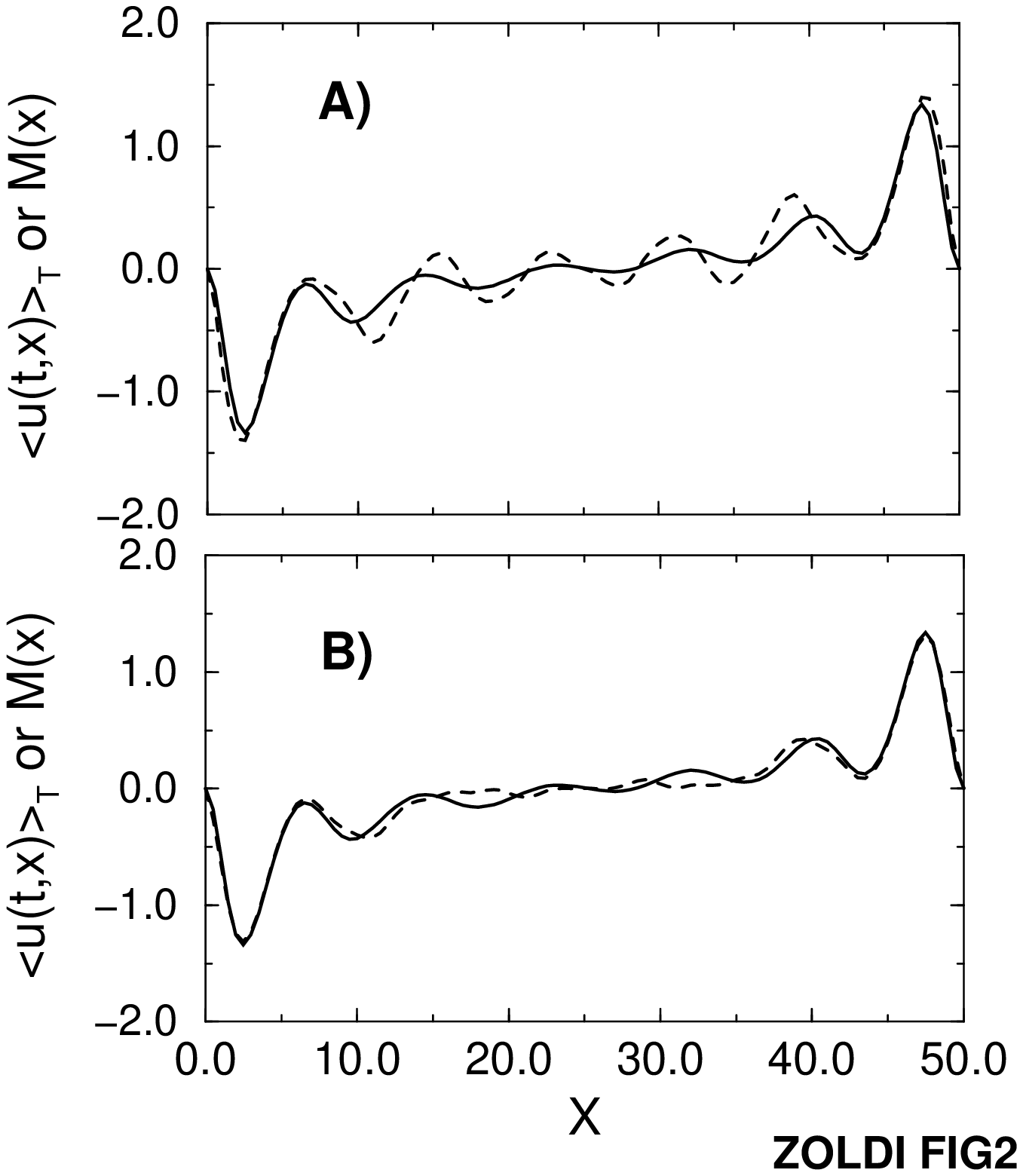}}

\newpage
\centerline{\epsfsize=25.5in \epsfbox{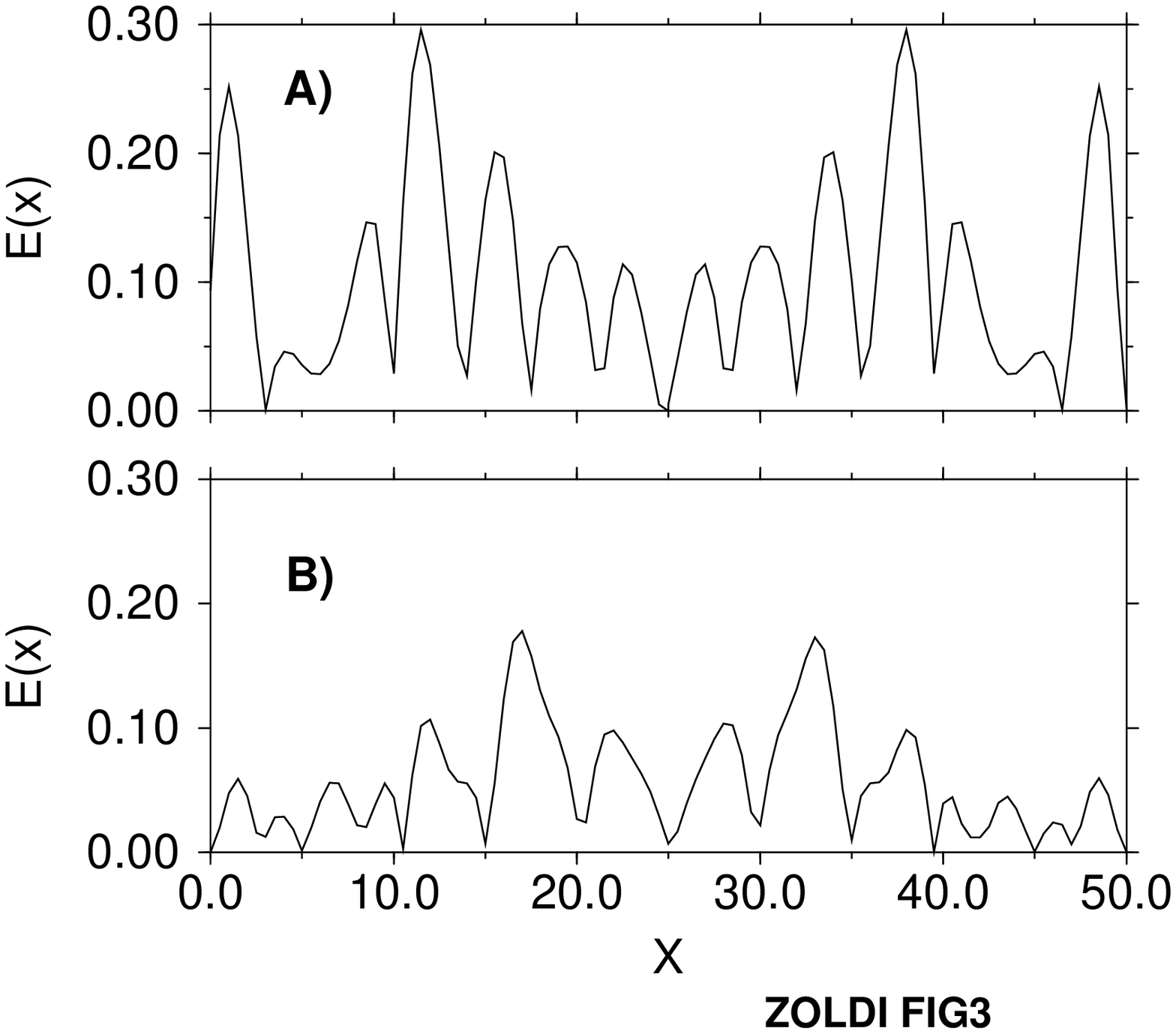}}

\end{document}